\documentclass[10pt, conference, letterpaper]{IEEEtran}
\IEEEoverridecommandlockouts
\usepackage{cite}
\usepackage{amsmath,amssymb,amsfonts}
\usepackage{algorithmic}
\usepackage{graphicx,subcaption}
\usepackage{textcomp}
\usepackage{xcolor}
\usepackage{xspace}
\usepackage{multirow}
\usepackage{booktabs}
\usepackage[hidelinks]{hyperref}
\usepackage{layouts} 

\def\BibTeX{{\rm B\kern-.05em{\sc i\kern-.025em b}\kern-.08em
    T\kern-.1667em\lower.7ex\hbox{E}\kern-.125emX}}
    
\usepackage{array}
\newcolumntype{L}[1]{
  >{\raggedright\let\newline\\\arraybackslash\hspace{0pt}}m{#1}}
\newcolumntype{C}[1]{
  >{\centering\let\newline\\\arraybackslash\hspace{0pt}}m{#1}}
\newcolumntype{R}[1]{
  >{\raggedleft\let\newline\\\arraybackslash\hspace{0pt}}m{#1}}

\newcommand{\iot}{IoT\xspace}
\newcommand{\pcap}{\texttt{pcap}\xspace}

\newcommand\fscore{$F_1$ score\xspace}
\newcommand\fig[1]{Figure~\ref{fig:#1}\xspace}

\newcommand{\one}{({\em i})\xspace}
\newcommand{\two}{({\em ii})\xspace}
\newcommand{\three}{({\em iii})\xspace}

\newcommand{\eg}{{\it e.g.,}\xspace}
\newcommand{\ie}{{\it i.e.,}\xspace}
\newcommand{\etal}{{\it et al.}\xspace}

\newcommand\s[1]{(\S\ref{s:#1})\xspace}

\IEEEoverridecommandlockouts \IEEEpubid{\makebox[\columnwidth]{978-3-903176-40-9~\copyright2021 Crown \hfill}\hspace{\columnsep}\makebox[\columnwidth]{ }}

\begin{document}

\title{Revisiting IoT Device Identification}

\author{\IEEEauthorblockN{Roman Kolcun}
\IEEEauthorblockA{\textit{University of Cambridge} \\
roman.kolcun@cl.cam.ac.uk}
\and
\IEEEauthorblockN{Diana Andreea Popescu}
\IEEEauthorblockA{\textit{University of Cambridge} \\
diana.popescu@cl.cam.ac.uk}
\and
\IEEEauthorblockN{Vadim Safronov}
\IEEEauthorblockA{\textit{University of Cambridge} \\
vadim.safronov@cl.cam.ac.uk}
\and
\IEEEauthorblockN{Poonam Yadav}
\IEEEauthorblockA{\textit{University of York} \\
poonam.yadav@york.ac.uk}
\and
\IEEEauthorblockN{Anna Maria Mandalari}
\IEEEauthorblockA{\textit{Imperial College London} \\
anna-maria.mandalari@imperial.ac.uk}
\and
\IEEEauthorblockN{Richard Mortier}
\IEEEauthorblockA{\textit{University of Cambridge} \\
richard.mortier@cl.cam.ac.uk}
\and
\IEEEauthorblockN{Hamed Haddadi}
\IEEEauthorblockA{\textit{Imperial College London} \\
h.haddadi@imperial.ac.uk}
}

\maketitle

\begin{abstract}

Internet-of-Things (IoT) devices are known to be the source of many security
problems, and as such, they would greatly benefit from automated management.
This requires robustly identifying devices so that appropriate network security
policies can be applied. We address this challenge by exploring how to
accurately identify IoT devices based on their network behavior, while
leveraging approaches previously proposed by other researchers. 

We compare the accuracy of four different previously proposed machine learning
models (tree-based and neural network-based) for identifying IoT device. We use
packet trace data collected over a period of six months from a large IoT
test-bed. We show that, while all models achieve high accuracy when evaluated on
the same dataset as they were trained on, their accuracy degrades over time,
when evaluated on data collected outside the training set. We show that on
average the models' accuracy degrades after a couple of weeks by up to 40
percentage points (on average between 12 and 21 percentage points).  We argue
that, in order to keep the models' accuracy at a high level, these need to be
continuously updated.  \end{abstract}


\section{Introduction}

Internet-of-Things (\iot) devices are the source of many security threats,
particularly in domestic deployments~\cite{Alrawi2019}. To counter such threats,
these devices would benefit from active and, in particular, automated
management. However, automating such tasks requires robustly identifying devices
to be able to apply appropriate policies, actions, and updates. In this
environment, the natural way of identifying \iot devices is to analyze their
network behavior at the home router: devices cannot hide behavior as, by
definition, they must interact over the network in order to provide
functionality.  Performing analyses of network behavior at the home router is
robust in terms of privacy, scalability and not relying on dependencies from
manufacturer-provided cloud-services. Furthermore, there is already nascent
support for summarizing such analyses’ results via the MUD
standard~\cite{rfc:8520}.

Previous work has resorted to machine learning to carry out \iot device
identification. The usual approach entails training machine learning models
offline or in a cloud environment~\cite{Miettinen2017, Hafeez2020,
Sivanathan2018, Pashamokhtari2020}, and run inference to identify the devices at
the home routers. However, the training and validation of these models is done
on a particular set of devices, and for a limited time period, thus achieving
high accuracy. These models' accuracy may drop when evaluated on different \iot
datasets or require continuous retraining to maintain the level of accuracy
needed. 

To investigate it in more detail, we evaluate three \iot device identification
approaches from the literature on the same dataset: \one a two-stage Random
Forest classifier using features extracted from a 1 hour window of collected
traffic \cite{Sivanathan2018}, \two a 2D Convolutional Neural Network on a
stream of raw packets \cite{Yin2021, Lopez-Martin2017}; and \three Random Forest
and Decision Tree classifiers on features extracted from 1 second window of
network traffic \cite{Pinheiro2019}. We also propose new \iot device
classification models, a Random Forest classifier and a Fully Connected Neural
Network trained on features extracted from TCP/UDP flows. We evaluate in total
six algorithms using four sets of features. We train and evaluate them on a 27
week-long dataset that we gathered from a large \iot test-bed comprising 41 \iot
devices, split into three time periods for training, while the evaluation was
done on the whole period. We show that \iot device identification models have
high accuracy only when the training and inference is run on the same dataset.
Our findings prove that static, pre-trained models cannot be used for
identification across different home \iot networks while ensuring high accuracy.
Thus, we conclude that it is paramount to update the models at the edge with new
incoming data.

The main contributions of the paper are as follows:
\begin{itemize}
    \item We gather a large measurement dataset of 27 weeks from a large \iot
    test-bed containing 41 \iot devices.
    \item We study and compare six different machine learning models for \iot
    device classification in terms of their accuracy using the dataset for
    training and inference.
    \item We show that in all six cases the accuracy decays over time,
    demonstrating the need for updating models at the edge.
    \item We release the extracted features (pre-processed \iot data) used to
    train the machine learning algorithms in order to further research in this
    area in the
    community\footnote{\url{https://github.com/DADABox/revisiting-iot-device-identification}}.
\end{itemize}

\section{Related Work} \label{s:related}

In the last decades, a vast number of machine learning-based network monitoring
and Internet traffic classification techniques, both in a distributed and
centralized manner, have been explored~\cite{Moore2005, Pacheco2019,
Nguyen2008}. However, not all methods are suitable for \iot, and some of these
techniques are adopted and customized for \iot; therefore, in this section, we
focus only on techniques used for analyzing \iot traffic. 

\textbf{Traffic Classification for \iot.} Offline \iot network traffic analysis
is used for understanding various \iot device or user behaviors~\cite{ren-imc19,
Apthorpe2016, Tahaei2020}. For example, Yadav~\etal~\cite{Yadav2019} studied
traffic from a dozen \iot devices in a lab environment to understand network
service (\eg DNS, NTP) dependencies and robustness of device function when
connectivity is disrupted.  Apthorpe~\etal~\cite{Apthorpe2016} analyzed the
traffic rates of four \iot devices, showing that observations about user
behavior can be inferred even from encrypted traffic. Similarly, traffic
categorization using both statistical and machine learning techniques has been
performed by Amar~\etal~\cite{Yousef2018}.  

\textbf{Device Identification and Anomaly detection in \iot.} The \iot device
identification is the first step towards finding any malicious or unknown \iot
device in the network. Generally, many \iot devices have a unique identifier
assigned during manufacturing such as MAC address or hardware serial numbers.
Even though these unique addresses could reveal some information about the
device manufacturer, still the full identification of malicious/abnormal devices
in the network using only these unique addresses is not possible. Thus,
behavior-based \iot device identification methods, which use traffic
classification mechanisms have gained attention recently~\cite{Meidan2017,
Miettinen2017, Hafeez2020, Trimananda2020, Saidi2020}. The \iot applications,
\eg anomaly detection and prediction, require low latency and privacy at the
edge, and traffic based behavior identification is needed to be done in the
real-time at the gateway level for security and data privacy
purpose~\cite{Yang2019, Magid2019, Kusupati2018}.  

\textbf{Machine Learning for Device Identification.}  Machine Learning in \iot
at the edge is still in its infancy, due to partly lack of available network
data in the wild and lack of compact machine learning models. The recent uptake
in resource-constrained machine learning~\cite{tflite,banbury2020, Painsky2019,
Feraudo2020} has led to a renewed interest in applying machine learning to \iot
network-related problems, specifically network traffic
classification~\cite{Ortiz2019, chimera2014, Feng2018}, anomaly
detection~\cite{Nguyen2019,Feraudo2020} and device
identification~\cite{Meidan2017, Miettinen2017, Hafeez2020, Sivanathan2018,
Kolcun2020}.  Sivanathan \etal~\cite{Sivanathan2018} used multi-stage
classifiers~(Naive Bayes Multinomial and Random Forest Classifier (RFC)) for
\iot device classification and achieved accuracy from 99.28\% to 99.76\% with
classifiers trained on 1 to 16 days data from 28 unique \iot devices. The high
accuracy achieved by this algorithm makes it the first choice for evaluation in
our work.

Nguyen \etal\cite{Nguyen2019} trained Gated Recurrent Network (GRU) for
federated learning  for anomaly detection using 33 devices categorized in 27
categories and for evaluation, deployed 13 devices and found only 5 are
vulnerable to the Mirai attack when the attack is injected in the local network.
The attack was detected within 30 minutes. Many identification works train
machine learning models offline or in a cloud environment~\cite{Miettinen2017,
Hafeez2020, Sivanathan2018, Pashamokhtari2020} and run inference to identify
\iot devices on local gateways. The training and evaluation is done only for a
set of devices for a limited time period, thus inference achieves a good
accuracy when testing data is similar to the training data. However, for real
world scenarios, a pre-trained model on a small set of devices would not work on
a large set of unknown \iot devices. The identification accuracy may drop when
the inference data is different from the training dataset, therefore, requiring
retraining of the model for the local setup. None of the works above have looked
at or addressed this problem; we are not only investigate retraining
requirements for maintaining high identification accuracy over an extended
period but also investigate the comparative performance of the \iot
identification algorithms on the same dataset.

\section{Dataset and Models} \label{s:models}

\begin{table}[]
\footnotesize
  \captionsetup{skip=0.2em, font=small}
  \caption{\label{table:devices}
    Categorized \iot devices in our test-bed.
    }

    \begin{tabular}{L{0.25\linewidth} L{0.65\linewidth}}
      \toprule
      \bf Category & \bf Device Name\\
      \midrule
      Surveillance & Blink camera, Bosiwo camera, D-link camera, Reolink camera
      Ring doorbell, UBell doorbell, Wansview camera, Yi camera, LeFun camera, ICSee doorbell\\
      \midrule
      Media           & Apple TV, Fire TV, Roku TV, LG TV, Samsung TV\\
      \midrule
      Audio           & Allure speaker, Echodot, Echospot, Echoplus, Google home\\
      \midrule
      Hub             & Insteon hub, Lightify hub, Philips hub, 
      Smartthings hub, Xiaomi hub, Switchbot hub, Blink security hub\\
      \midrule
      Appliance       & Smart Kettle, Smarter coffee machine, Sousvide cooker, Xiaomi rice cooker\\
      \midrule
      Home automation & Honeywell thermostat, Nest thermostat, Netatmo weather
      station, TP-link bulb, TP-link plug, Wemo plug, Xiaomi plug, Smartlife
      remote, Smartlife bulb, Meross door opener\\
      \bottomrule
    \end{tabular}
\end{table}

\subsection{Dataset}
\label{s:dataset}

To capture data, we built a test-bed that currently comprises 41 different \iot
devices. Table~\ref{table:devices} describes the devices in our test-beds, by
category. In each category, these devices were chosen as the most popular one by
a large online retailer.

In the test-bed, in addition to the devices, a Linux server running Ubuntu 18.04
with two Wi-Fi cards for 2.4~GHz and 5~GHz connections, plus two 1~Gbps Ethernet
connections for LAN and Internet connectivity are part of the setup.  The server
sits outside of any firewall and has a public IPv4 address. However, to match a
regular home network environment, all \iot devices are behind a NAT setup and
cannot be accessed directly from the Internet. The monitoring software
automatically detects the connection of a new device to the network, assigns it
a local IP address, and starts capturing packets using \emph{tcpdump}. Each
device's traffic is filtered by MAC address into separate files. Each file is
stored in a \texttt{pcap} format. 

Each device is assigned a unique \emph{device ID}. In some cases, we
extract features from TCP/UDP flows. Each flow is identified by a 5-tuple
\texttt{(source IP address, source port, destination IP address, destination
port, transport protocol)}. Each device generates multiple flows. When a flow is
extracted from the \texttt{pcap} file, it is assigned a \emph{device ID}. Each
flow is classified independently of the other flows generated by the same or
other devices.

Data were collected over a period of 27 weeks. During this period the
interaction with the devices was rather sporadic. Given that the devices were
deployed in the lab with people present during the daytime, some of the devices
might have been occasionally activated, \eg cameras with motion detection or a
smart speaker reacting to the user's request. However, these activations were
rather rare. Researchers did not perform any software update that required
manual intervention, \ie confirming the update using the accompanying app.
However, some devices might have updated themselves without researchers'
knowledge (\eg Amazon Alexa devices update themselves automatically).
Researchers also sporadically checked whether the devices are still connected to
the Internet and functioning properly. This check has been done via a cell phone
using an accompanying app.

We are aware that the test-bed setup is not an approximation of a usual
household environment. In the household environment the interaction with the
devices could be expected to occur more often and probably follow some pattern.
In our test-bed scenario, these devices were mostly idle. Additionally, our
test-bed contains \iot devices only, \ie non-\iot devices such as laptops,
computers, or cell phones are not present. The only cell phones connected to the
test-bed are used to control the \iot devices. The traffic generated by these
cell phones are stored in separate \texttt{pcap} files and are not included in
the training dataset.

To evaluate whether the model can reliably classify \iot devices, we split the
collected data into three periods, each 9 weeks long. We then train each ML
model using data from only one period, while evaluating them on the whole
dataset of 27 weeks.

\begin{center}
\begin{table}
\footnotesize
 \captionsetup{skip=0.2em, font=small}
\caption{\label{tab:unsw-features}
List of features extracted from 1 hour window of network traffic used for
training of two-phase RFC models.}
\begin{tabular}{l l}
\toprule
\bf Feature Name         & \bf Feature Description  \\
\toprule
{\it First Stage}   & Using Naive Bayes Multinomial Classifier    \\
\midrule
\-\hspace{0.2cm} {\it bag\_of\_ports} & list of ports contacted \\
\-\hspace{0.2cm} {\it bag\_of\_domains} & list of domains contacted \\
\-\hspace{0.2cm} {\it bag\_of\_ciphers} & list of used cipher suites \\
\midrule
{\it Second Stage}  & Using Random Forest Classifier \\
\midrule
\-\hspace{0.2cm} {\it flow\_volume} & volume of flow \\

\-\hspace{0.2cm} {\it flow\_duration} & duration of flow \\
\-\hspace{0.2cm} {\it flow\_rate} & rate of flow \\
\-\hspace{0.2cm} {\it sleep\_time} & sleep time \\
\-\hspace{0.2cm} {\it dns\_interval} & interval between DNS requests  \\
\-\hspace{0.2cm} {\it ntp\_interval} & interval between NTP requests \\
\-\hspace{0.2cm} {\it ports\_class} & class for bag of ports ($1^{st}$ stage) \\
\-\hspace{0.2cm} {\it ports\_confidence} & confidence for bag of ports ($1^{st}$ stage) \\
\-\hspace{0.2cm} {\it domain\_class} & class for bag of domains ($1^{st}$ stage) \\
\-\hspace{0.2cm} {\it domain\_confidence} & confidence for bag of domain ($1^{st}$ stage) \\
\-\hspace{0.2cm} {\it cipher\_class} & class for bag of cipher suites ($1^{st}$ stage) \\
\-\hspace{0.2cm} {\it cipher\_confidence} & confidence for bag of cipher suites ($1^{st}$ stage) \\
\bottomrule

\end{tabular}
\end{table}
\end{center}

\begin{center}
\begin{table}
\footnotesize
 \captionsetup{skip=0.2em, font=small}
\caption{\label{tab:brazil-features}
List of features extracted from 1 second window of network traffic for training
of RFC and DTC models.}
\begin{tabular}{l l}
\toprule
\bf Feature Name         & \bf Feature Description  \\
\midrule
{\it bytes\_sum}    & sum of bytes of all packets                 \\   
{\it bytes\_avg}    & average size of packets                     \\
{\it bytes\_std}    & standard deviation of size of packets       \\
\bottomrule

\end{tabular}
\end{table}
\end{center}

\begin{center}
\begin{table}
\footnotesize
 \captionsetup{skip=0.2em, font=small}
\caption{\label{tab:features}
List of features extracted from TCP/UDP flows used for training of RFC and FC-NN models.}
\begin{tabular}{l l}
\toprule
\bf Feature Name         & \bf Feature Description  \\
\midrule
{\it src\_port}     & source port                                 \\   
{\it dest\_port}    & destination port                            \\
{\it bytes\_out}    & number of bytes sent                        \\
{\it bytes\_in}     & number of bytes received                    \\
{\it pkts\_out}     & number of packets sent                      \\
{\it pkts\_in}      & number of packets received                  \\
{\it ipt\_mean}     & mean of inter-packet interval               \\
{\it ipt\_std}      & standart deviation of inter-packet interval \\  
{\it ipt\_var}      & variance of inter-packet interval           \\
{\it ipt\_skew}     & skewness of inter-packet interval           \\
{\it ipt\_kurtosis} & kurtosis of inter-packet interval           \\
{\it b\_mean}       & mean of packet sizes                        \\
{\it b\_std}        & standard deviation of packet sizes          \\
{\it b\_var}        & variance of packet sizes                    \\
{\it b\_skew}       & skewness of packet sizes                    \\
{\it b\_kurtosis}   & kurtosis of packet sizes                    \\
{\it duration}      & duration of the stream                      \\
{\it protocol}      & protocol ID                                 \\
{\it domain}        & second and top level domain                 \\ 
\bottomrule

\end{tabular}
\end{table}
\end{center}

\subsection{Processing Traces}\label{s:traces}

In this subsection, we list four different  \iot device identification
models that we analyzed and describe how data were pre-processed for each of them.

\subsubsection{Multi-stage RFC with 1 Hour Window}

This is one of the pioneering work in \iot device identification by Sivanathan
\etal~\cite{Sivanathan2018}. This classifier extracts features from a traffic
collected over the duration of one hour. This two-stage approach uses the
combination of Naive Bayes Multinomial (NBM) classifier and the Random Forest
classifier (RFC). In the first stage, NBM is used to classify bag of words, \ie
contacted domain names, port numbers, and cipher suites used. In the second,
stage the output from the first stage and the following features from the 1 hour
window of traffic are extracted: flow volume, flow duration, flow rate, sleep
time, DNS interval, and NTP interval. These features are summarized in
Table~\ref{tab:unsw-features}. After processing the captured traffic, our
dataset contains 130,460 records.

\subsubsection{2D Convolutional Neural Networks on Raw Packet Data}

Convolutional networks (CNN) were used in image classification for a very long
time. Their advantage is that they are capable of automatic extracting of
features. Researchers have been using them to process raw packets to either
classify an \iot device \cite{Yin2021} or a network flow \cite{Lopez-Martin2017}. 
For each TCP or UDP flow first $X$ packets are order in rows.
Then for each packet first $Y$ bytes are extracted. If the packet is smaller
than $Y$ or there are fewer than $X$ packets in the flow, the empty space is padded
with zeroes. Additionally, fields from the IP header which can uniquely
identify the device, such as source MAC or IP address, are replaced with
zeroes. This yields an input grid of the size $X \times Y$ which is passed to the
CNN. In this paper we used $X = 10$ and $Y = 250$ \cite{Yin2021}. After
processing all TCP/UDP flows, our dataset
consists of 19,771,368 input grids. 

The evaluated convolutional network consisted of the following layers:
Convolutional, MaxPooling, Convolutional, MaxPooling, Flatten, Dropout, and an
output Dense layer. 

\subsubsection{Multiple Classifiers on 1 Second Window}

In this case researchers extracted three features from each second of traffic
generated by an \iot device \cite{Pinheiro2019}. These features are the sum of the
size of all packets, the average size of a packet, and the standard deviation of
the size of a packet. These features are summarized in
Table~\ref{tab:brazil-features}. The researchers then evaluated five different
classifiers: Random Forest, Decision Tree, Support Vector Classifier, K-Nearest
Neighbors, and Voting Classifier. The Voting Classifier used four previous
classifiers to classify the 1 second network trace by choosing the device with
most votes. After processing the captured traffic, our dataset contains
75,396,781 records.

\subsubsection{RFC and FC-NN on TCP/UDP Flows}

We also propose a new \iot device classification system based on features
extracted from TCP and UDP flows. We processed each \pcap file using
\emph{joy}\footnote{\url{https://github.com/cisco/joy}} utility which extracts the
following features from each TCP/UDP network flow (summarized in
Table~\ref{tab:features}): source and destination IP address, source and
destination port number, number of packets sent and received, bytes of packets
sent and received, starting and ending time of the flow. Additionally,
\emph{joy} extracts DNS request and replies which can be later analyzed. Flow
features are extracted if the network flow is inactive for more than ten
seconds, or if the network flow is active for more than 30 seconds. If the
network flow continues, a new record is created. It means, that a set of
features is extracted at latest after 30 seconds.

The extracted features contain also information about the first up to \emph{N}
packets. We used the default value of $N = 50$. This information includes data
about packet sizes and inter-packet intervals. Using information about packets,
additional features are computed, \ie duration of the flow, and for both, packet
sizes and inter-packet intervals, mean, standard deviation, variance, skew, and
kurtosis is computed. Each flow is assigned the \emph{device ID}. 

The list of DNS responses is used to map IP addresses to domain names. We chose
not to use IP addresses as a feature because they may not be consistent due to
the nature of the services running in cloud. A virtual server may migrate to
another physical server and its IP may change. Or a new server might be
temporarily started to balance the load. Additionally, many large manufactures
are using DNS load balancing where the same domain is translated to different IP
addresses. Therefore, we decided to use the domain name as a feature. However,
we noticed that many times the domain name differs on the third or further
level. This is especially common when a content delivery network is contacted.
Therefore, we decided to use only the second and top level domain name as a
feature. In our dataset we identified only 153 unique second and top level
domain names. The final dataset contains 60,653,581 records. 

The dataset is evaluated using Random Forest classifier and Fully Connected
neural network. The fully connected network consists of a Dense input layer, two
hidden Dense layers, and an output Dense layer.

\section{Evaluation}\label{s:evaluation}

In this section we evaluate the selected machine learning models on various
data. The dataset spans over the period of 27 weeks. The dataset is split into
three equally-sized periods covering weeks 1-9, 10-18, and 19-26 respectively.
Each model is trained on one period using stratified sampling with 80\%-20\% split
between the training and testing set. For the evaluation, the whole dataset
spanning 27 weeks is split into 1-week chunks and each week is evaluated as a whole.

Machine learning based classifiers (NBM, RFC, and DTC) are implemented using
python scikit-learn library\footnote{\url{https://scikit-learn.org/}}. All classifiers
were using default settings. Neural network based classifiers were implemented
using Keras library\footnote{\url{https://keras.io/}}. All models were trained for the
duration of 50 epochs and the model with the highest accuracy was chosen for the
evaluation.

Data pre-processing and feature extraction is described in \s{traces}. 
All models were evaluated on a server using two Intel Xeon Gold 6132 CPU @
2.6~GHz with 256~GB of RAM running Ubuntu~18.04 operating system. 
 
We used a standard evaluation metric~\cite{Hackeling2014}, \fscore for the
overall measure of the models accuracy and is defined as:

\[ Precision = \frac{True\: Positive}{True\: Positive + False\: Positive} \]

\[ Recall = \frac{True\: Positive}{True\: Positive + False\: Negative} \]

\[ F_1 = 2 \times \frac{Precision \times Recall}{Precision + Recall}, F_1 \in
\langle 0, 1 \rangle \]

where $True\: Positive$ represents the number of times when a device was
correctly classified, $False\: Positive$ the number of times when the device was
classified as some other device, $True\: Negative$ the number of times when the
device is correctly not classified, and $False\: Negative$ the number of times
when a device is classified instead of a correct device. \fscore represents a
harmonic mean of precision and recall.

\begin{figure*}[!bpt]
    \centering
    \begin{subfigure}[t]{.49\linewidth}
\includegraphics[width=\linewidth]{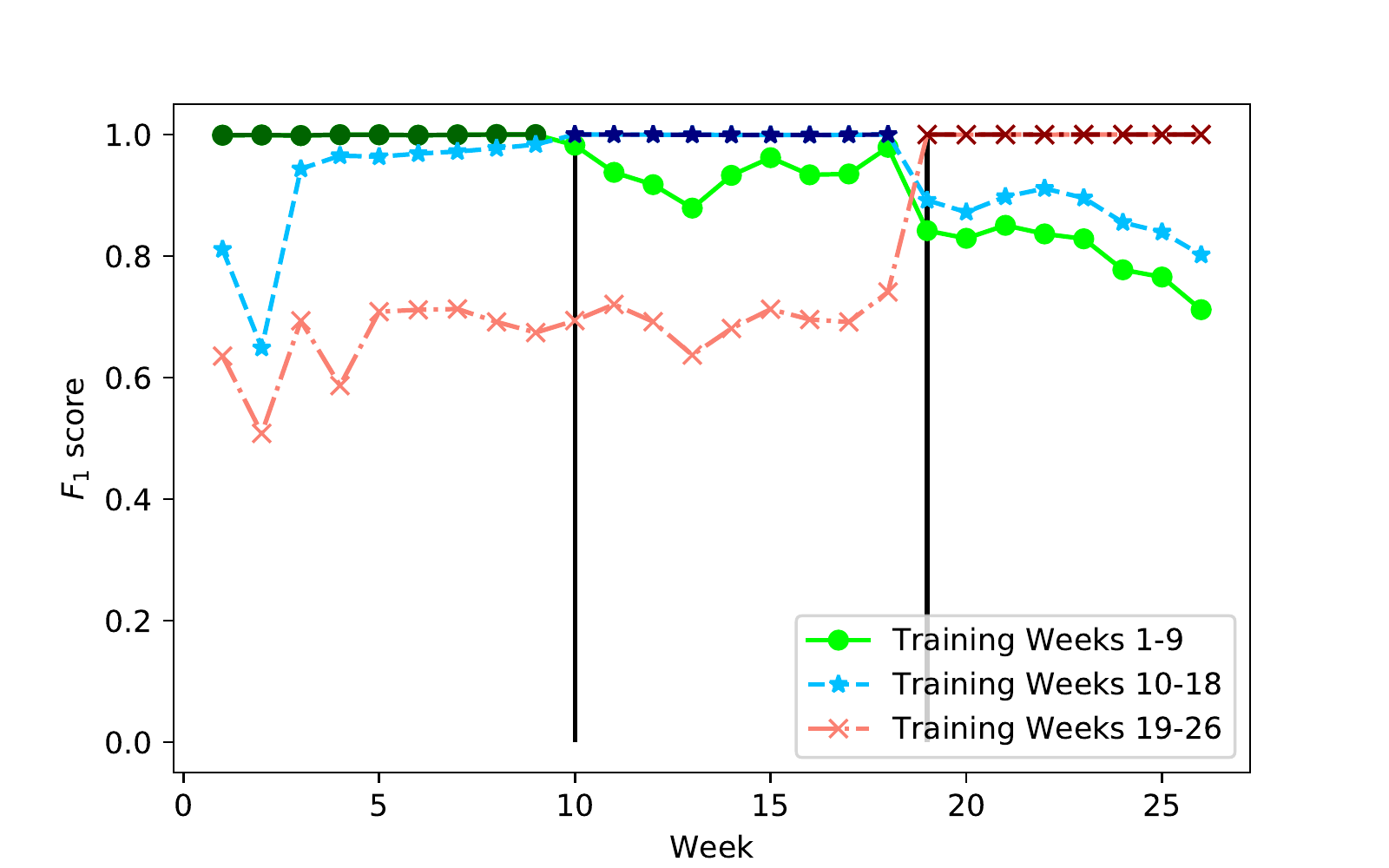}
    \caption{1 hour window with 2-stage NBM + RFC}
    \label{fig:unsw}
    \end{subfigure}
    \begin{subfigure}[t]{.49\linewidth}
\includegraphics[width=\linewidth]{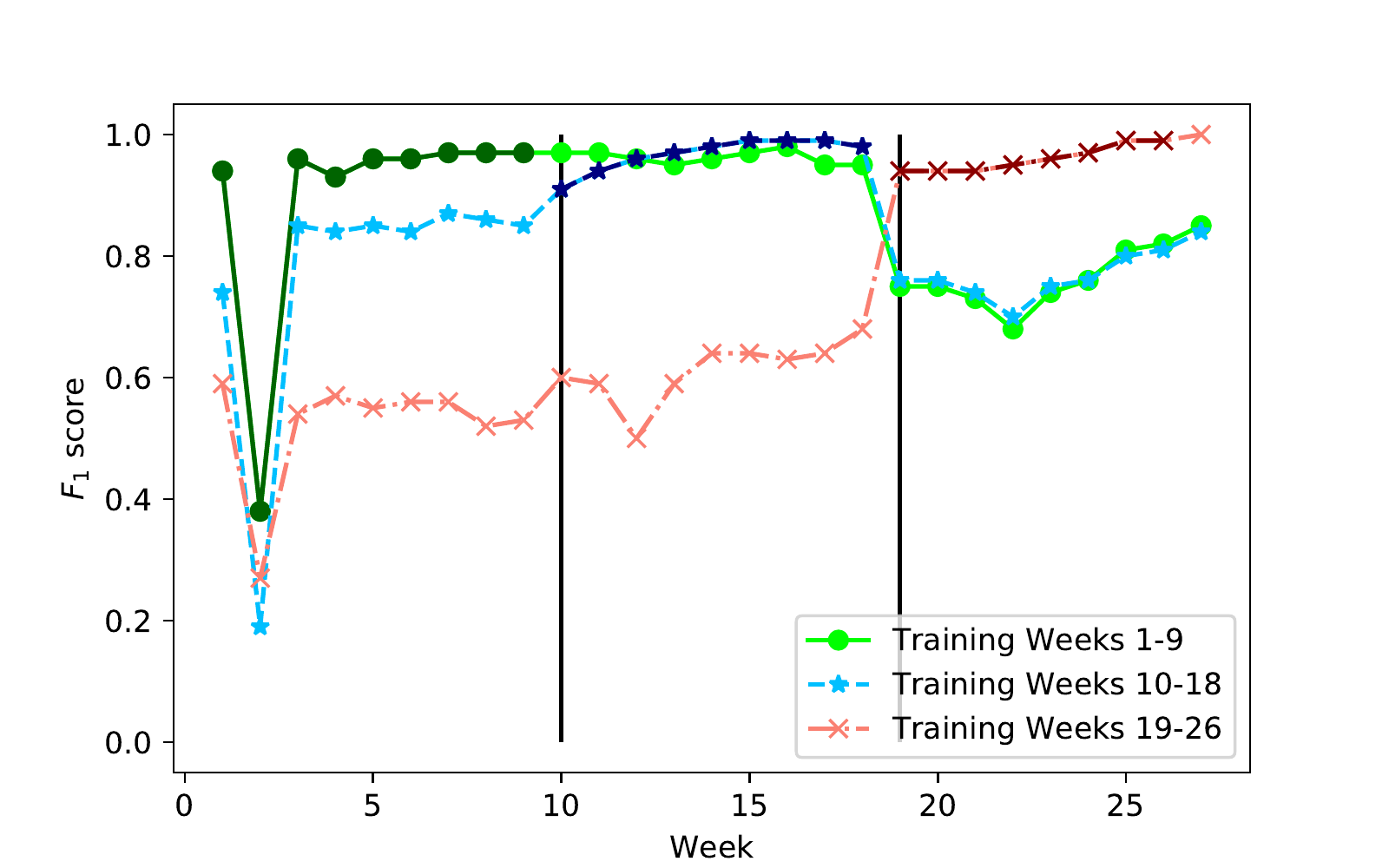}
    \caption{Raw packets with 2D Convolutional Network}
    \label{fig:etei}
    \end{subfigure}
    \begin{subfigure}[t]{.49\linewidth}
\includegraphics[width=\linewidth]{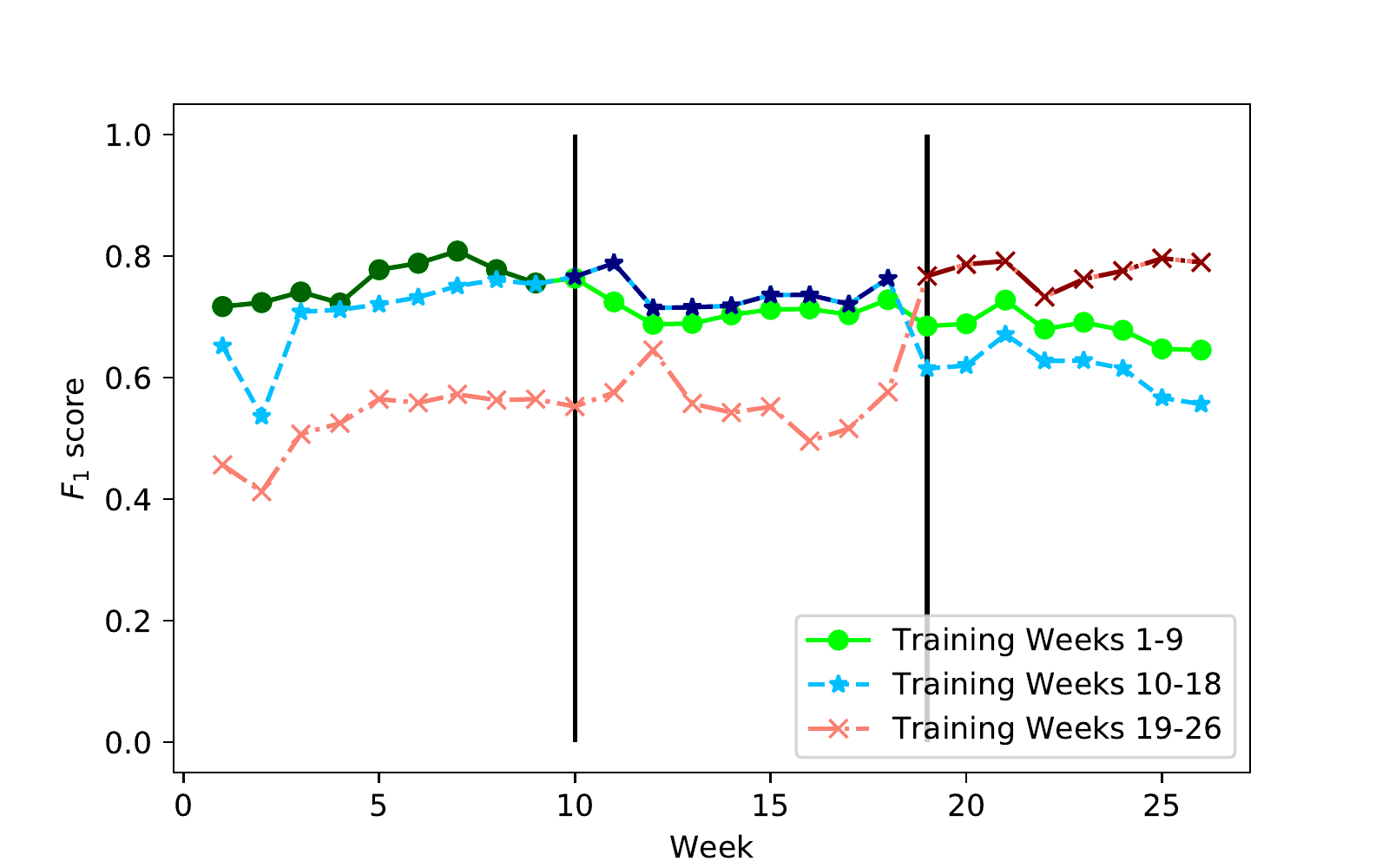}
    \caption{1 second window with RFC}
    \label{fig:brazil-rfc}
    \end{subfigure}
    \begin{subfigure}[t]{.49\linewidth}
\includegraphics[width=\linewidth]{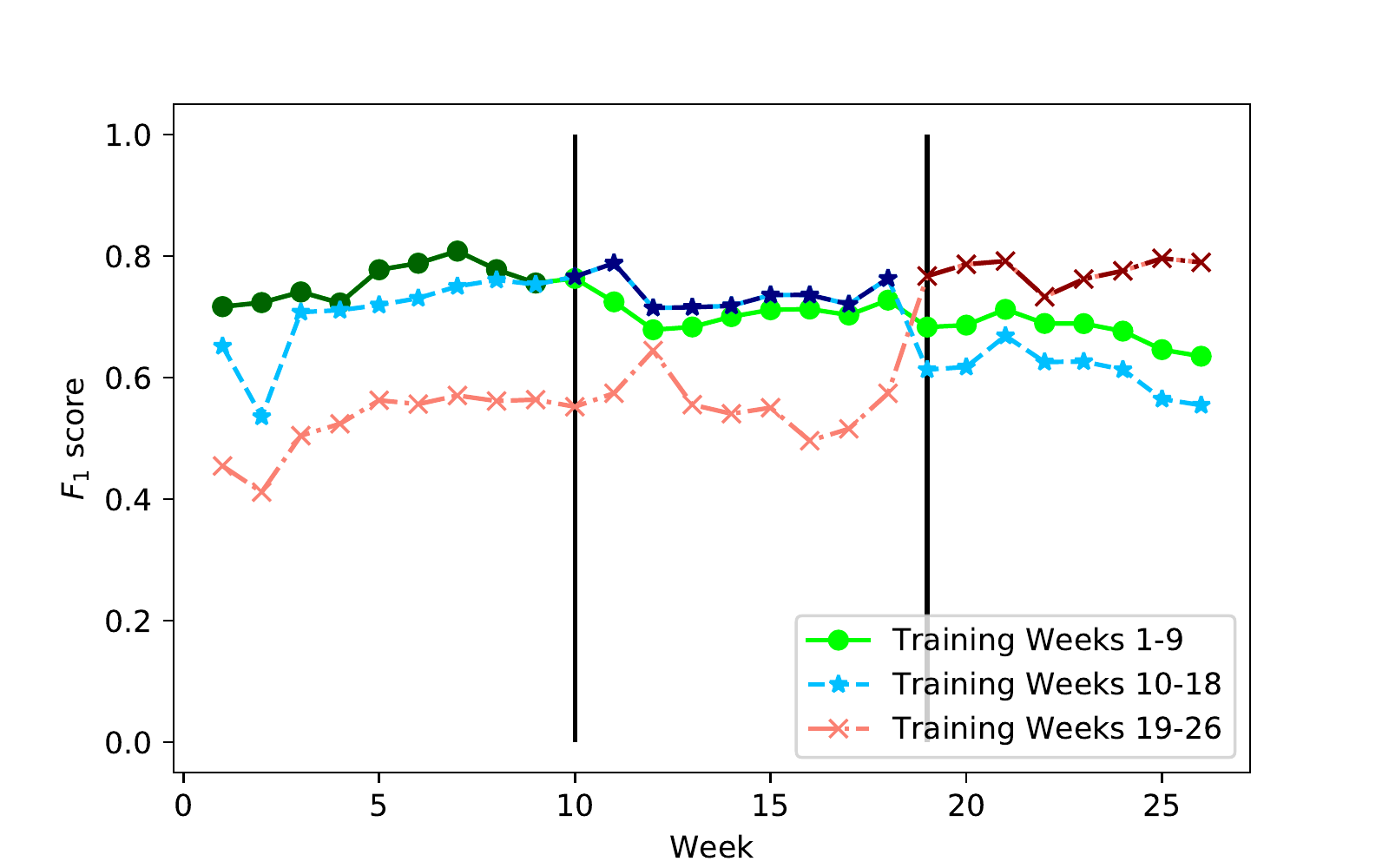}
    \caption{1 second window with DTC}
    \label{fig:brazil-dtc}
    \end{subfigure}
    \begin{subfigure}[t]{.49\linewidth}
\includegraphics[width=\linewidth]{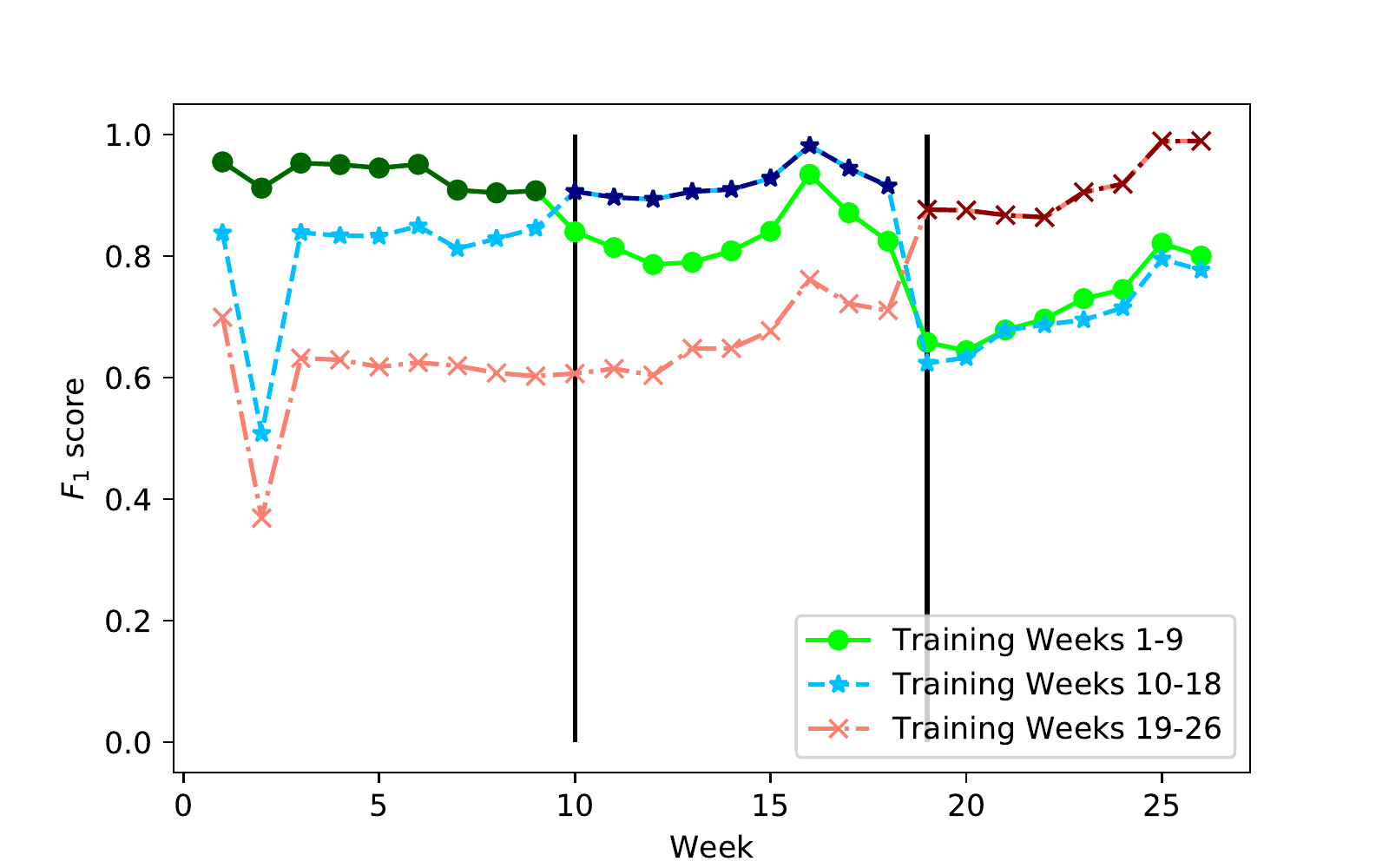}
    \caption{TCP/UDP flows with RFC}
    \label{fig:feat-rfc}
    \end{subfigure}
    \begin{subfigure}[t]{.49\linewidth}
\includegraphics[width=\linewidth]{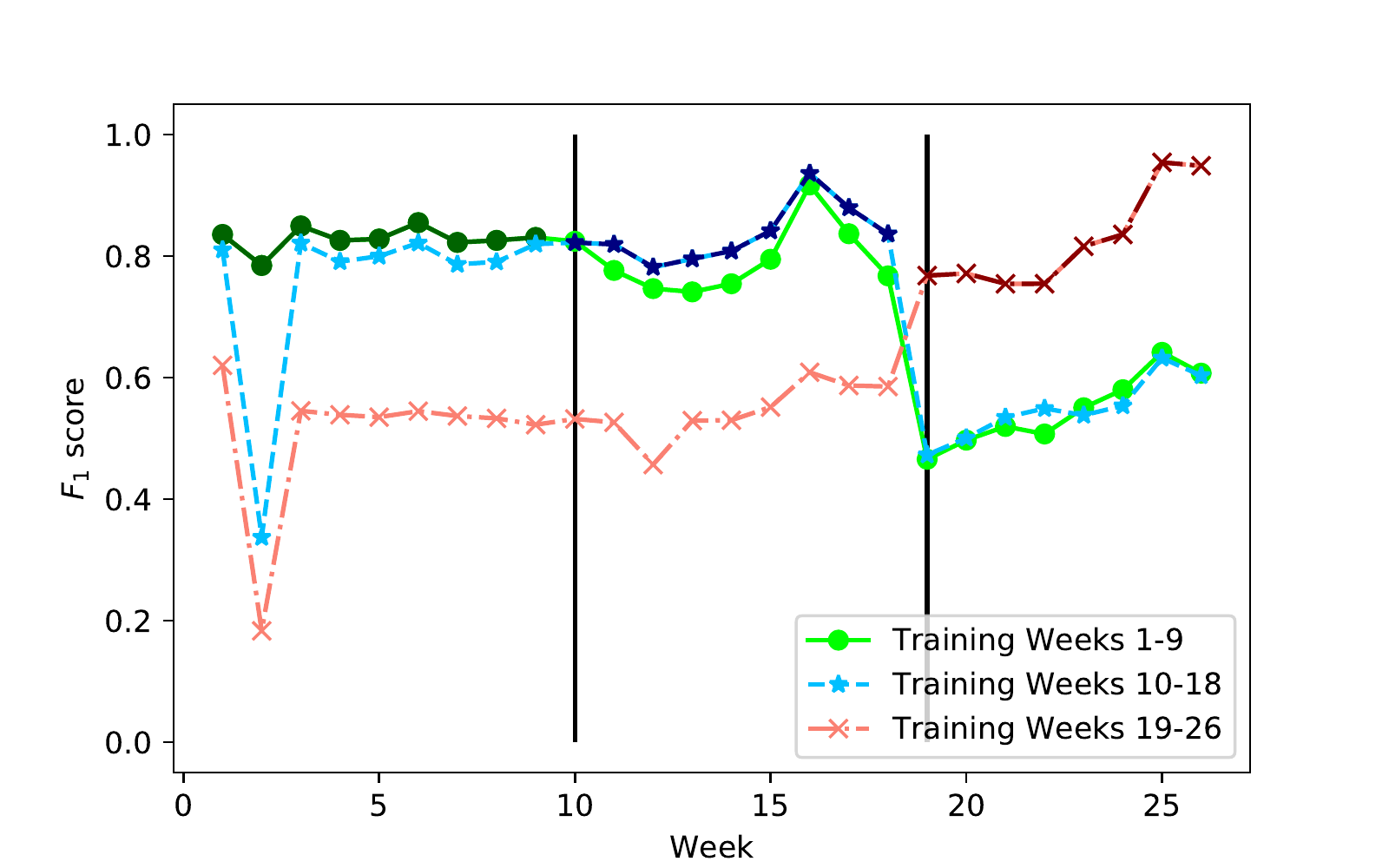}
    \caption{TCP/UDP flows with Fully Connected NN}
    \label{fig:feat-nn}
    \end{subfigure}

    \caption{\fscore of various classifiers on various datasets. Each line
    corresponds to a different training window (\ie weeks 1-9, 10-18, or 19-26).
    The darker color shows the period when the model is evaluated on the
    training data. Vertical bars splits the figures into three separate periods.}
    \label{fig:classifiers}
\end{figure*}

\subsection{Multi-stage RFC with 1 Hour Window}

The model relies on two stage evaluation and is evaluated on statistical data
collected over the period of 1 hour \cite{Sivanathan2018}. The evaluation of
this model can be seen in \fig{unsw}. This model achieves the near-perfect
\fscore when evaluated on the same data as the training set. Even if the dataset
is split into 80\%-20\% between training and testing data, because the testing
data is chosen from the same time period as the training data, the model
achieves very high \fscore.

Because the features are extracted from a network traffic over a longer period
of time (\ie 1 hour), the model can achieve higher accuracy for devices with a
relative small and regular network footprint, \eg a weather station. However,
devices which rely more on the interaction with a user or its environment, such
as smart speakers or doorbell cameras with a motion detection, can produce
various amount of traffic during different parts of a day. Additionally, this
approach relies on knowing all the contacted domains, port numbers, and used
cipher suites in advance. This knowledge cannot always be obtained in advance as
a firmware update might change any or all of these variables. Additionally,
domain names might be automatically generated by a load balancer.

However, when the same model is evaluated on a dataset on another time period than the
training period, the accuracy of the model degrades with the absolute time
difference between the training and the testing dataset. This applies to all
three training periods. On average, the \fscore decreases by 18 percentage
points when comparing the performance of models evaluated on the training period
vs. other periods. 

\subsection{2D Convolutional Neural Networks on Raw Packet Data}

Several researchers have used 2D Convolutional Neural Networks (CNN) to
automatically extract features and perform classification of the \iot devices or
network flows \cite{Yin2021, Lopez-Martin2017}. These approaches are based on
ordering the packets into a grid of a fixed size and passing it to the 2D CNN.
Usually, these multiple CNN layers are followed by other layers which can
slightly increase the accuracy of classification. Because of the computational
complexity and time restrictions, we evaluated a simpler model consisting of 2D
CNN layers only. However, we believe that the results can be extrapolated to
other more complex models. 

\fig{etei} shows performance of the CNN in \iot device classification. Similar
to the two-stage RFC, the best \fscore is achieved when evaluated on the dataset
from the same period as the training dataset. Surprisingly, the accuracy is
rather high and is very close to the two-stage RFC. Additionally, CNN can
achieve such high \fscore using only a small subset of packets from each flow,
as opposed to collecting and analyzing traffic from an hour long window. CNN
also do not require to know any information in advance (\eg contacted domain
names) or manual extraction of features. The results show that CNNs are a
promising technique for \iot device classification. 

However, similar to other approaches, the high accuracy is achieved only when
evaluated on the same period as the training set. Outside of the training
period, the accuracy gradually degrades, even though not as quickly as with
other models. The average difference in the \fscore between the models evaluated
on the training dataset and other periods is 21 percentage points. 

The models were trained for 50 epochs with batch size of 128. They achieved
their peak accuracy very close to the $50^{th}$ epoch.

\subsection{Multiple Classifiers on 1 Second Window}

Authors of this model showed that even a simple statistics on a 1 second window
worth of network traffic can achieve rather high classification accuracy
\cite{Pinheiro2019}. The advantage of this approach is its simplicity, very low
computational and memory overhead, and the fact that it does not need to know
any prior information (\eg contacted domain names). The researchers showed, that
RFC is the best single model, however, by combining results from multiple
different machine learning models, it is possible to achieve even higher
accuracy. In this case they combined four different ML models: RFC, DTC, SVC,
and K-NN. Even though we tried to replicate these results, because our dataset
was incomparably larger than the one used by authors, we were not able to finish
the training of the SVC and K-NN models. Therefore, we present here the results
using the RFC (\fig{brazil-rfc}) and the DTC (\fig{brazil-dtc}) models only. 

Figures show that the \fscore achieved by both models is virtually the same.
However, it can be seen the similar trend as previously. The highest accuracy
models achieve when they are evaluated on the data from the same period as they
were trained on. However, when they are evaluated on the dataset outside their
training period, their accuracy degrades over time. This is most visible with
the model trained on the last period. When this model is evaluated on first two
periods, it achieves significantly lower \fscore. On average, the \fscore is
lower by 12 percentage points in the case of RFC and 13 percentage points in the
case of DTC. 

\subsection{RFC and FC-NN on TCP/UDP Flows}

The last evaluated approach is based on TCP/UDP flows. We have trained two
classifiers on these datasets: RFC (\fig{feat-rfc}) and the Fully Connected NN
(\fig{feat-nn}). As can be seen in the figures, the overall performance of both
models is rather similar, with the difference that RFC performs on average by 7
percentage points better than the NN model. As expected, also these two models
show similar \fscore degradation when evaluated outside their training period.
On average, the RFC model achieves \fscore 20 percentage point lower when
compared to evaluation on the training period, while the \fscore of Fully
Connected NN is lower on average by 21 percentage points.

Interestingly, the size of the RFC models ranged between 100-150 GB which made
it challenging to evaluate. Even though the neural network models were trained
for 50 epochs, they achieved their peak accuracy no later than in $11^{th}$
epoch. Since then, the accuracy was constantly decreasing and by the $50^{th}$
epoch it was more than 10 percentage points lower.

\section{Discussion}\label{s:discussion}

\fig{classifiers} shows that no matter which classifier is used, either based on
classical machine learning techniques or on neural networks, they all lose their
accuracy over time. While their accuracy is reasonably high when evaluated on
the data from the same time period as they were trained on, once they are used
for evaluation on data outside of this period, their accuracy decreases with the
absolute difference in time. 

The same applies to the set of extracted features (or the lack of them). Whether
the features are extracted from an hour long window with some prior knowledge of
data, a simple one second window statistics, a set of raw bytes from packets, or
an elaborate features extracted from TCP/UDP flows - all of these features lead
to accuracy degradation when evaluated outside of the training dataset. 

What is also important is that the data produced for this experiment are from an
academic test-bed deployed in a rather controlled environment with a very small
interaction of researchers with the test-bed. Thus, we can assume that data
produced by various sets of devices in users' home will vary much more and
therefore the models might degrade even faster. 

Because the dataset was obtained from a test-bed in a controlled environment,
the network traces contain mostly background traffic of these devices. In a home
environment with a regular interaction with the devices the number of different
network flows will increase, which will lead to potentially larger search space
and lower accuracy.

Interesting behavior can be seen when comparing the \fscore of \emph{all} models
outside their training period. This fact is the most visible with the models
trained on data from period one and two (green and blue line, respectively) and
evaluated on data from period three. For \emph{all} models and \emph{all}
various features, these two lines are almost parallel. Similar behavior can also
be seen with red and blue lines in period one and red and green line in period
two. This fact suggests that the network traffic produced by the devices changes
over time and is not a problem of a particular feature extraction or a machine
learning model. 

Another interesting data point is the second week where the \fscore of most of
the models suddenly drops. However, if the data are included in the training set
(dark green line), most of the models are capable to learn this anomaly and
perform with high accuracy. This has one obvious exception - 2D Convolutional
Neural Network (\fig{etei}) whose \fscore drops to less than 40\%. Again,
because this behavior is consistent across various models and features, it can
be assumed that the pattern of the network traffic of the devices has
significantly changed.

Because the similar decrease in classification accuracy over time can be seen
across different machine learning models and using different set of features, it
suggests that the problem of \iot device identification is not a problem of a
single approach, but rather a bigger problem that requires further research. It
appears that the network traffic generated by these devices changes over time
and therefore a single model cannot stay accurate for a longer period of time.
In this paper, we have not investigated the root cause of the accuracy decline.
We have not found any obvious reason for this behavior and change in network
traffic. We plan to defer the further investigation for the future work. We also
plan to investigate whether certain class of devices is more susceptible to
change of network traffic patterns than the others. 

In this paper we have used out-of-the-box parameters for ML models and we did
not do any fine-tuning of the parameters. Fine-tuning of the parameters is a
rather time consuming activity that would also require significant computational
resources. We believe, that fine-tuning of the parameters would lead to
marginally higher accuracy, but it would not affect the trend of decreasing
accuracy over time.

It can also be expected that because these devices also communicate among
themselves, a model] created for a device in one network setting might not be
accurate for the same device in another network setting. Additionally, users'
usage pattern change between households and therefore it can be expected that a
device might have a significantly different network traffic pattern, even when
it is deployed in the same network setting.

In order to study this scenario, we are working on a home router device, that
will allow users to deploy small test-beds in their homes. The user would
connect their \iot devices to the router which would collect the traffic
generated by these devices. The router would send anonymised packet headers
(without payloads) to the server, where we would be able to analyze them.

One of the possible solutions that we would like to investigate in the future is
to keep updating the model with new data at the edge, meaning retraining or
tuning of models on edge devices \cite{federatedGoogle, paulik2021federated}.
This could potentially not only solve the problem of changing traffic pattern
over time, but also the problem of a device deployed in different network
settings.

\section{Conclusion}\label{s:conclusion}

In this paper we revisited four different approaches to \iot device
identification based on the network traffic. For that purpose we have collected
network traffic from a test-bed containing 41 different \iot devices over the
period of 27 weeks. We have split these data into three periods, each containing
9 non-overlapping weeks. Then we have used three various approaches found in the
literature. First, we evaluated a two-stage Random Forest classifier using
features extracted from a 1 hour window of collected traffic
\cite{Sivanathan2018}. Next, we used 2D Convolutional Neural Network on a stream
of raw packets \cite{Yin2021, Lopez-Martin2017}. Later, we used Random Forest
and Decision Tree classifiers on features extracted from a 1 second window of
network traffic \cite{Pinheiro2019}. Finally, we proposed and evaluated models
using Random Forest classifier and Fully Connected Neural Network on features
extracted from TCP/UDP flows. 

Each model was trained using data from one period, while it was evaluated on data
from the whole dataset (\ie all three periods). We have shown that while the
accuracy of these models is high when tested on the dataset from the same period
as the training dataset, the accuracy degrades over time when evaluated on
dataset collected outside of the training period. The average degradation of the
models' accuracy ranged between 12 and 21 percentage points, with an average of
17 percentage points. 

Because this behavior is consistent across all models and various features
extracted, we believe that the data generated by the devices change over time
and cannot be captured by a single model. We propose a possible solution for
this problem is by updating the model at the edge \cite{federatedGoogle,
paulik2021federated}.

\section*{Acknowledgment}

We would like to thank Pinheiro \etal \cite{Pinheiro2019} for providing us with
their source code so the evaluation of their approach could be fair and
complete. We thank the anonymous reviewers and our shepherd, Maciej Korczynski.
Authors were partially funded by EPSRC DADA: Defence Against Dark Artefacts
(EP/R03351X/1) and EPSRC Databox: Privacy-Aware Infrastructure for Managing
Personal Data (EP/N028260/1). 

\bibliographystyle{IEEEtran}
\bibliography{main}

\end{document}